\documentclass[
journal=ancac3, 
manuscript=article]{achemso}

\usepackage{siunitx}
\usepackage[english=british]{csquotes}
\usepackage{xcolor}

\usepackage{xr}

\makeatletter
\newcommand*{\addFileDependency}[1]{
  \typeout{(#1)}
  \@addtofilelist{#1}
  \IfFileExists{#1}{}{\typeout{No file #1.}}
}
\makeatother
 
\newcommand*{\myexternaldocument}[1]{%
    \externaldocument{#1}%
    \addFileDependency{#1.tex}%
    \addFileDependency{#1.aux}%
}

\myexternaldocument{mainSI}

\DeclareSIUnit\Molar{\textsc{m}}

\author{Thomas Kister}
\affiliation[INM --- Leibniz Institute for New Materials]
{INM --- Leibniz Institute for New Materials, Campus D2 2, 66123 Saarbr\"ucken, Germany}

\author{Debora Monego}
\affiliation[University of Sydney]
{ARC Centre of Excellence in Exciton Science, School of Chemistry, University of Sydney, Sydney, New South Wales 2006, Australia}

\author{Paul Mulvaney}
\affiliation[University of Melbourne]
{ARC Centre of Excellence in Exciton Science, School of Chemistry, University of Melbourne, Parkville, Victoria 3010, Australia}

\author{Asaph Widmer-Cooper}
\affiliation[University of Sydney]
{ARC Centre of Excellence in Exciton Science, School of Chemistry, University of Sydney, Sydney, New South Wales 2006, Australia}

\author{Tobias Kraus}
\email{tobias.kraus@leibniz-inm.de}
\affiliation[INM --- Leibniz Institute for New Materials]
{INM --- Leibniz Institute for New Materials, Campus D2 2, 66123 Saarbr\"ucken, Germany\\
Colloid and interface chemistry, Saarland University, Campus D2 2, 66123 Saarbr\"ucken, Germany}

\title[\texttt{achemso} demonstration]
{On the Colloidal Stability of Apolar Nanoparticles:\\ The Role of Particle Size and Ligand Shell Structure}

\begin{document}

\begin{abstract}
Being able to predict and tune the colloidal stability of nanoparticles is essential for a wide range of applications, yet our ability to do so is currently poor due to a lack of understanding of how they interact with one another. Here, we show that the agglomeration of apolar particles is dominated by either the core or the ligand shell, depending on the particle size and materials. We do this by using Small-Angle X-ray Scattering and molecular dynamics simulations to characterize the interaction between hexadecanethiol passivated gold nanoparticles in decane solvent. For smaller particles, the agglomeration temperature and interparticle spacing are determined by ordering of the ligand shell into bundles of aligned ligands that attract one another and {interlock}. In contrast, the agglomeration of larger particles is driven by van der Waals attraction between the gold cores, which eventually becomes strong enough to compress the ligand shell. Our results provide a microscopic description of the forces that determine the colloidal stability of apolar nanoparticles and explain why classical colloid theory fails.
\end{abstract}

Following the advent of the hot-injection method for nanocrystal synthesis \cite{Murray1993}, there has been a plethora of studies on the preparation of an enormous range of materials in nanocrystal form including noble metals, magnetic materials such as Fe3O4 and FePt, quantum dots such as CdS and CdSe, upconverters including YLaF4, core-shell nanocrystals such as Au@Ag, ternary materials like CuInS2, perovskites, and alloys. In all these diverse systems, inorganic core particles are stabilized in an apolar solvent by a self-assembled layer of surfactant. Prototypical examples are metal and semiconductor nanocrystals or quantum dots \cite{badia1996self} with ligand shells of alkanethiols that are stable in organic solvents. Their applications include inkjet printed structures for detectors \cite{singh2010inkjet, jensen2011inkjet}, conductive inks and color-improving additives for OLEDs \cite{li2005white, talapin2010prospects}. Despite their widespread study, there is no generally accepted theory that explains when and why such systems will be colloidally stable, \textit{\textit{i.e.}} no extant model correctly predicts the colloidal stability of hydrocarbon capped inorganic nanoparticles dispersed in organic solvents. 

Existing approaches based on classical colloid theory describe the interaction between apolar nanoparticles by assuming a linear combination of contributions from dispersive van der Waals (vdW) attraction between the inorganic cores, the free energy of ligand/solvent demixing, and elastic energy due to deformation of the ligand shell \cite{khan2009self, goubet2011forces, sigman2004metal}. While the vdW attraction between the cores should be well described by Hamaker-Lifshitz theory, the model of the shell appears to be lacking. The free energy of ligand/solvent demixing is currently modelled using Flory-Huggins theory using Hildebrand solubility parameters \cite{flory1942thermodynamics} that were developed for molecular solutions, and assumes that the ligand/solvent interface can be adequately described by a density distribution that is radially symmetric and constant with temperature. The theory used to describe the elastic energy makes a similar assumption about how the ligands are distributed in space.

In contrast, there is substantial evidence from both simulation and experiment that the structure of the ligand shell depends on core size, temperature, and solvent quality. Linear ligands change their arrangement in space in response to a decrease in temperature or solvent quality, aligning with one another and packing together.\cite{Luedtke1998, Ghorai2007, Lane2010, ramin2011odd, bolintineanu2014effects, Korgel2002} This affects both the ligand and solvent density distributions, which can become highly asymmetric about small particles,\cite{Luedtke1998,Ghorai2007, Lane2010, Widmer-Cooper2014} and can be expected to result in deviations from the theory described above. Indeed, simulations have already shown that the interaction between particles can change rapidly from repulsive to attractive as the ligands order.\cite{Widmer-Cooper2014, Widmer-Cooper2016} The importance of such breakdowns from the assumptions of classical colloid theory, and some of the effects that they can have on the interaction between nanoparticles regardless of ligand and solvent polarity, has recently been highlighted.\cite{batista2015nonadditivity}

Here, we combine experiments and simulations to characterize the stability of hexa\-decane\-thiol-coated gold nanoparticles with core diameters of \SIrange{4}{10}{\nano\meter} in decane. Their agglomeration and interparticle spacing was characterized as a function of particle diameter and temperature using X-ray scattering and compared with molecular dynamics studies of the ligand order and the interaction potential between pairs of particles. We find two different regimes depending on particle size: for small particles, the agglomeration is driven by the ordering of the ligand shells, while for larger particles, the vdW attraction between the cores becomes strong enough to drive agglomeration before the ligands order. This transition from ligand- to core-dominated agglomeration results in a nonlinear change in the interparticle spacing as the size increases. In the ligand-dominated regime, the interparticle spacing increases with particle size as the ligand shell becomes denser and the gaps between the ordered bundles decrease in size until the spacing is roughly equal to twice the thickness of the ligand shell. Beyond this point, the core-core vdW attraction drives the transition, eventually leading to compression of the ligand shell and a decrease again in the particle spacing.

The temperature-dependent stability that we measured and simulated for our nanoparticle dispersions was irreconcilable with interaction potentials calculated using Khan's expressions \cite{khan2009self}, even when modifying the Flory-Huggins parameters. Classical colloid theory predicts colloidal stability near room temperature for the small particles used here, while our experiments show rapid agglomeration, in some cases well above room temperature. We conclude that improved models will need to account for temperature-dependent transitions in the ligand shell, including their potential to change the symmetry of the ligand distribution about the particles.

\begin{figure}[htpb]
    \centering
    \includegraphics[width=0.4\linewidth]{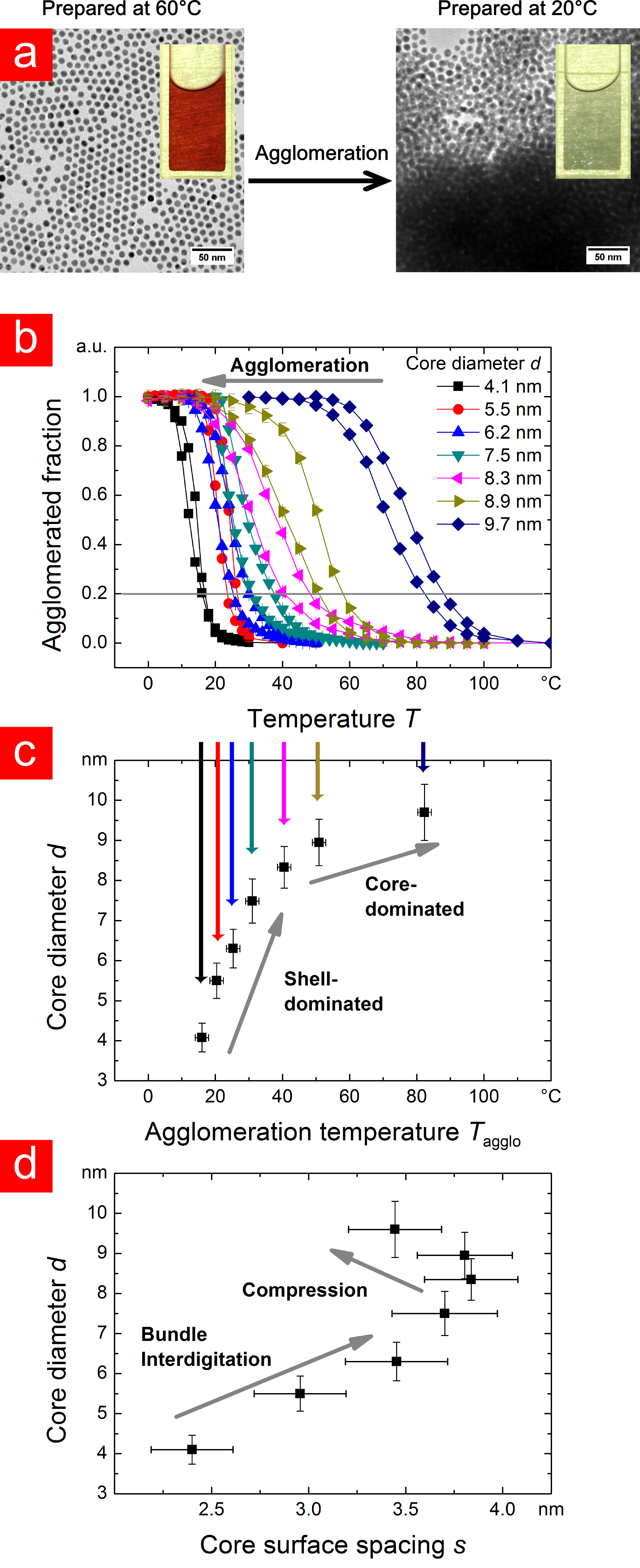}
    \caption{Temperature-dependant agglomeration of AuNP with 1-hexadecanethiol shells and different core diameters in decane. (a) Transmission electron micrographs of AuNP with a diameter of \SI{7.5}{\nano\meter}. The samples were prepared at \SI{60}{\celsius} and \SI{20}{\celsius}, respectively. Insets shows optical photographs of macroscopic dispersions at the respective temperature. (b) Fraction of agglomerated particles as determined by \emph{in situ} Small-Angle X-ray Scattering. All particles were dispersed at high temperatures, and agglomeration occurred upon cooling as indicated by the increase in structure factor. (c) Agglomeration temperature (where \SI{20}{\percent} of particles were agglomerated) as a function of core size. Note the change near a core diameter of  \SI{8.3}{\nano\meter}. (d) Core surface spacing between AuNP at the agglomeration temperature. The spacing is largest for the \SI{8.3}{\nano\meter} particles.}
    \label{Figure_1}
    \vspace{0.5em}
\end{figure}

\section{Results and Discussion}

Gold particles (AuNP) with 7 different mean core diameters between \SI{4}{\nano\meter} and \SI{9.7}{\nano\meter} and size distributions with widths below 10\% (mean diameter over standard deviation) were coated with 1-hexa\-decane\-thiol and dispersed in decane. Small angle X-ray scattering (SAXS) and transmission electron microscopy (TEM) data of the NPs are shown in Figures S1 and S2. The dispersions had concentrations of \SI{2.5}{\milli\gram\per\milli\liter} (\SI{0.013}{vol \percent}) and ligand shell densities of \SI{5.5}{\nano\meter^{-2}} (see Figure S3). Table \ref{table_size} provides a summary of all particles used.

The temperature-dependant colloidal stability was quantified using \emph{in situ} Small-Angle X-ray Scattering. Figure S4 shows how the data was analyzed. At high temperatures, all particles were dispersed, and the scattering corresponded to the form factor \(P(q)\) of dispersed spheres: \cite{schnablegger2013saxs} a Bessel function, as expected for spherical particles. Upon cooling, the particles agglomerated (see Figure S4a) and a peak appeared (for example, at \(q = \SI{0.594}{\nano\meter^{-1}}\) for AuNP with a diameter of \SI{9.7}{\nano\meter}, Figure S4b) \cite{pontoni2003microstructure, sztucki2006kinetic} due to the agglomerates' structure factor \(S(q)\). The peak height and area are directly proportional to the fraction of nanoparticles that agglomerated \cite{johnson1959x}. We fitted it with a Lorentz function to find the agglomeration temperature of the particles and the spacing between the particles (Figure S4c).

Figure \ref{Figure_1}a shows TEM micrographs of AuNP with a diameter of \SI{7.5}{\nano\meter} prepared at \SI{60}{\celsius} and \SI{20}{\celsius}. The insets shows the AuNP dispersion at the corresponding temperatures. A clear temperature dependent change is observable. Figure \ref{Figure_1}b shows how the fraction of agglomerated particles increased upon cooling for different core diameters. Smaller particles were more stable and agglomerated at lower temperatures than larger particles. The agglomeration temperature \(T_\mathrm{agglo}\), defined as the temperature at which \SI{20}{\percent} of particles had agglomerated, showed a strong nonlinear dependence on particle size, increasing rapidly beyond a diameter of roughly \SI{8}{\nano\meter} (Figure \ref{Figure_1}c). At the same time, the particle separation at \(T_\mathrm{agglo}\) reached a maximum around a diameter of \SI{8}{\nano\meter} before decreasing again (Figure \ref{Figure_1}d).

\begin{figure}[ht!]
    \centering
    \includegraphics[width=0.4\linewidth]{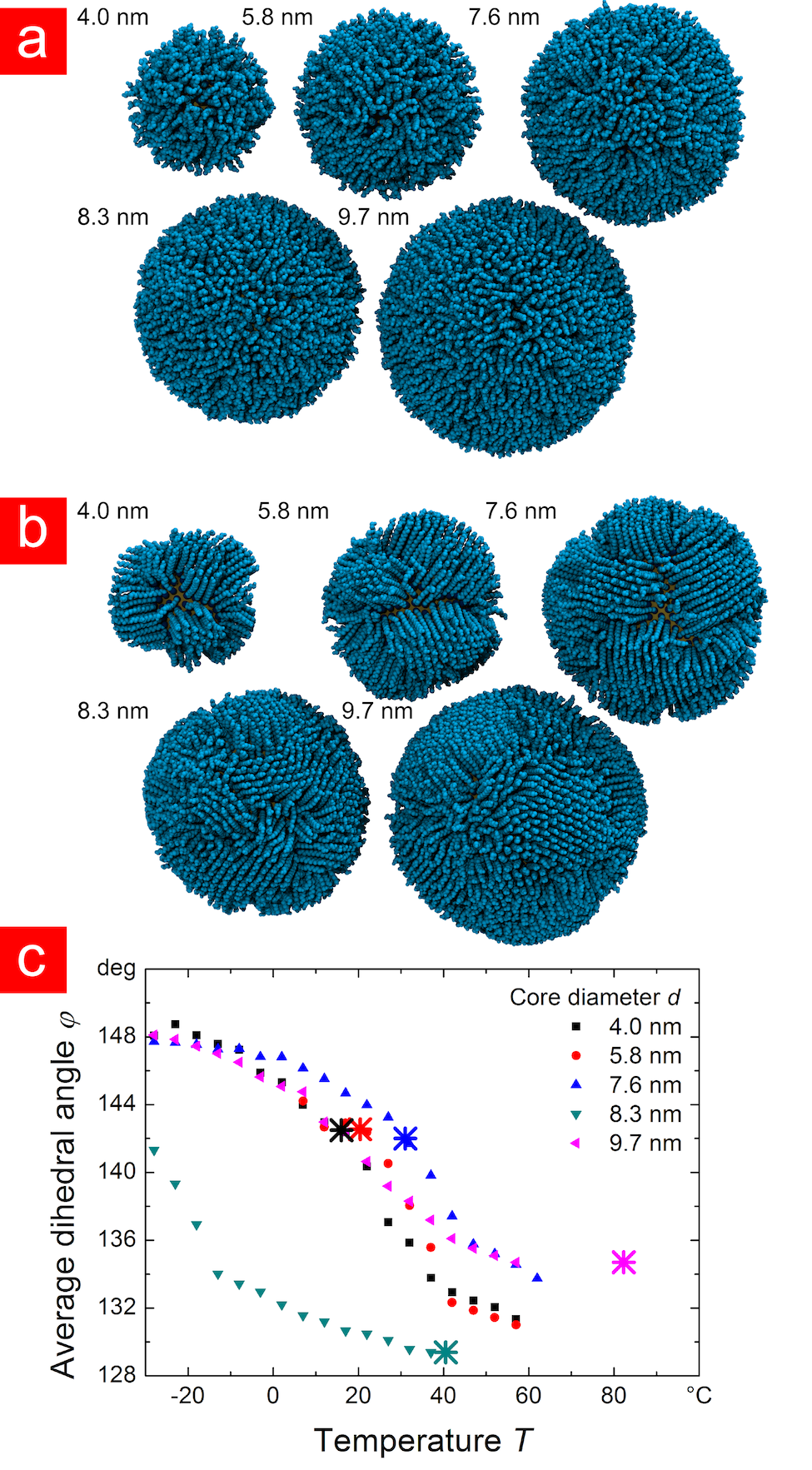}
    \caption{Simulation snapshots at (a) high and (b) low temperature show that the structure of the ligand shell depends on both temperature and particle size (decane solvent not shown). These changes are quantified by (c) the average dihedral angle of the ligands, which increases rapidly as the ligands order. For comparison, the experimental agglomeration temperatures have been indicated by large crossed symbols.}
    \label{Figure_2}
    \vspace{0.5em}
\end{figure}

We believe that the nonlinear, size-dependent stability of the nanoparticle colloids is due to a transition from ligand- to core-dominated agglomeration as the particle size increases. Large-scale molecular dynamics simulations of the nanoparticles in explicit n-decane solvent support this hypothesis. Snapshots show the structure of the ligand shell well above (Figure \ref{Figure_2}a) and below (Figure \ref{Figure_2}b) the agglomeration temperature. At high temperature, the ligands are mobile and the shell disordered. In contrast, at low temperature, the ligands are well-ordered and much less mobile, with the ligands adopting mostly all-trans conformations and aligning with one another. This causes the ligands to cluster into bundles and results in ligand shells that are increasingly anisotropic as the particle size decreases. Similar changes have been observed in simulations of other small spherical and rod-shaped nanoparticles as the temperature or solvent quality was reduced \cite{Luedtke1998, Lane2010, Widmer-Cooper2014}.

\begin{figure}[ht]
    \centering
    \includegraphics[width=0.4\linewidth]{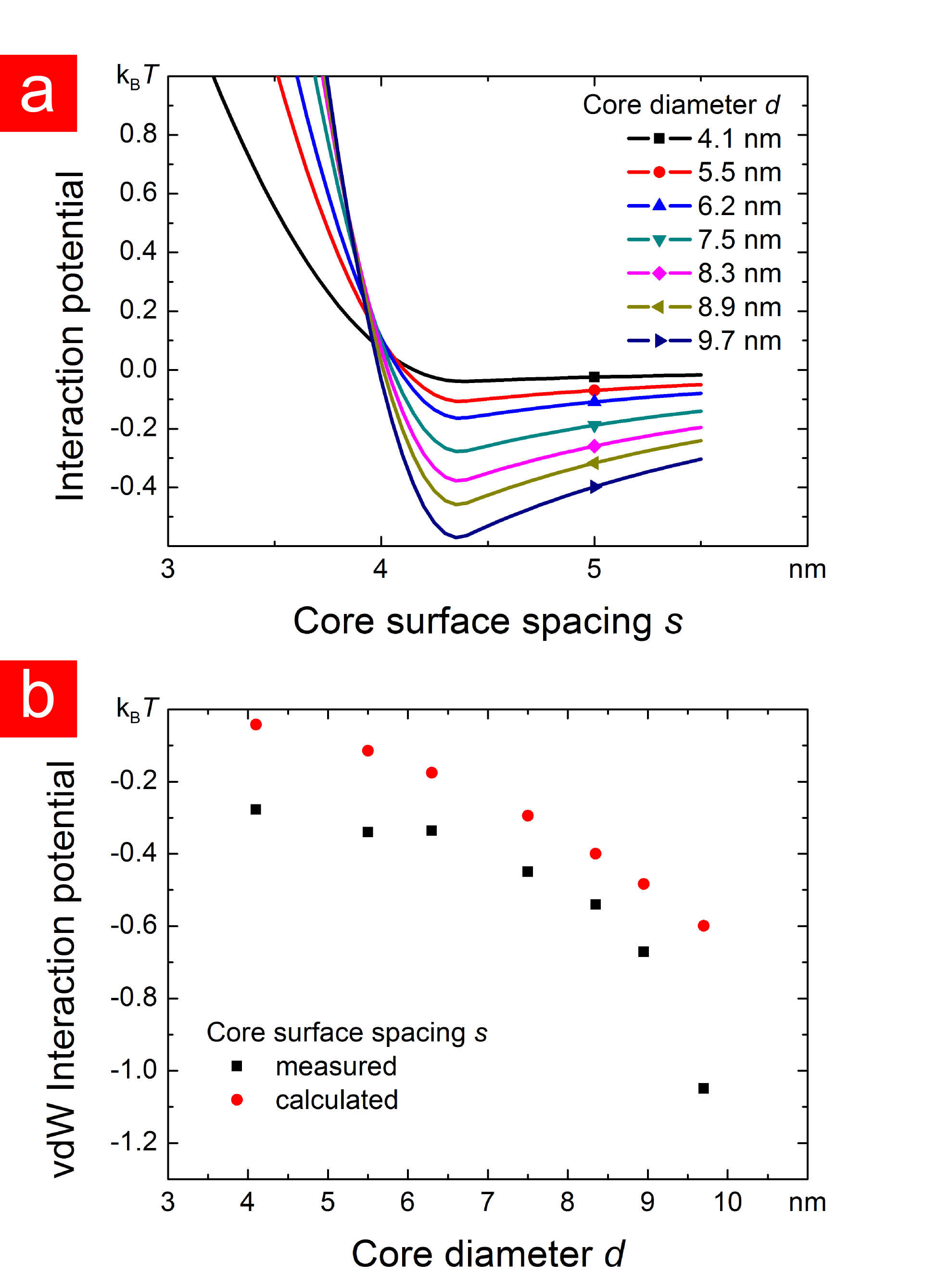}
    \caption{(a) Interaction potentials calculated using equations 1 to 4 using solubility parameters of the free ligands. The predicted potentials are incompatible with experiments and simulations. (b) Van der Waals interaction potentials between the Au cores at the experimental and the calculated particle separations in units of thermal energy at \(T_\mathrm{agglo}\) and \(s_\mathrm{agglo}\) as estimated using Hamaker-Lishitz theory.}
    \label{analytical}
    \vspace{1.5em}
\end{figure}

The degree of order of the ligand shell can be quantified using the dihedral angle of the ligand molecules in the shell. Figure \ref{Figure_2}c shows the average dihedral angle as a function of temperature for different core diameters. We define an \enquote{ordering temperature} \(T_\mathrm{order}\) at which the average dihedral angle equals \ang{140}. For particles from \SIrange{4}{7.6}{\nano\meter} in diameter, the ligand ordering (small symbols) preceded particle agglomeration (large crosses) and exhibited the same dependence on particle size, indicating that the agglomeration is driven by ordering of the ligand shell. This is supported by analytical calculations of the vdW interaction between small Au cores that turn out to be too weak to cause agglomeration at the experimentally observed particle separations (see Figure \ref{analytical}b below).

In contrast, larger particles agglomerated before the ligands had ordered, indicating that their agglomeration is driven {primarily} by attraction between the nanocrystal cores. Analytical calculations of the Au-Au vdW interaction are consistent with a shift to core-dominated agglomeration, with the Au-Au interaction predicted to exceed \(0.5 \mathrm{k_B} T\) at the experimental interparticle spacing at \(T_\mathrm{agglo}\) around a diameter of \SI{8}{\nano\meter}. In agglomerates, each particle will have an average of roughly 12 neighbours, which at \(0.5 k_\mathrm{B} T\) per interaction would result in a total stabilization energy per particle of \(6 \times 0.5 = 3 \mathrm{k_B} T\). In comparison, Lennard-Jones particles, which have a slightly longer relative interaction range, are known to aggregate around an interaction energy of \(0.75 \mathrm{k_B} T\) per particle pair \cite{Johnson1993}. The sudden drop in \(T_\mathrm{order}\) above a diameter of \SI{8}{\nano\meter} is not essential for this conclusion, but the much smaller drop in vacuum (see Figure S5) indicates that it is due to a significant change in how the solvent interacts with the ligands as the diameter increases. The radial probability distributions for the ligand and solvent (see Figure S7) show that the solvent becomes increasingly excluded from the ligand shell, especially in the disordered state. 

\begin{figure}[ht!]
    \centering
    \includegraphics[width=0.4\linewidth]{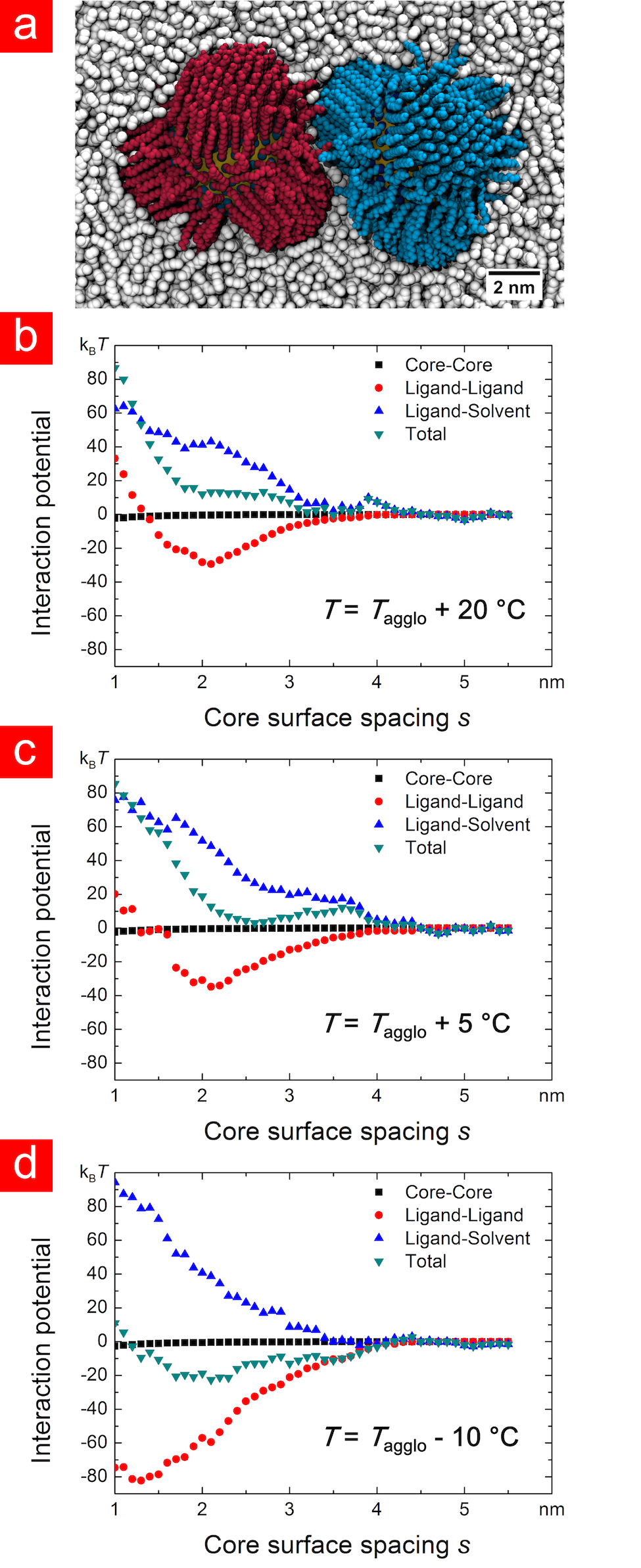}
    \caption{(a) Simulation snapshot at \(T_\mathrm{agglo} + \SI{5}{\celsius}\) showing two \SI{4}{\nano\meter} particles at \SI{2.4}{\nano\meter} separation. Different colors have been used for the ligands on the two particles and the solvent partially hidden. Interaction potentials for pairs of \SI{4}{\nano\meter} Au particles in decane at (b) \(T_\mathrm{agglo}\) + \SI{20}{\celsius}, (c) \(T_\mathrm{agglo} + \SI{5}{\celsius}\), and (d) \(T_\mathrm{agglo}\) \SI{- 10}{\celsius}, obtained from constrained molecular dynamics simulations. The total potential is green, the ligand-ligand contribution is red, the ligand-solvent contribution is blue, and the core-core contribution is black.}
    \label{PMF}
    \vspace{0.5em}
\end{figure}

Our simulations explain a second unexpected experimental result: the non-linear relation between core size and core surface spacing shown in Figure \ref{Figure_1}d. We used Ehrenfest's equation to calculate the spacing of particles in amorphous agglomerates and Bragg's equation for the spacing in crystalline agglomerates \cite{guinier1963x}. The resulting surface spacings (Figure \ref{Figure_1}d) were remarkably well-defined and reproducible. The spacings increased with particle core size up to a diameter of \SI{8.3}{\nano\meter} and decreased for larger cores, an effect that is readily explained by the molecular shell structure seen in simulation: Small particles have large surface curvatures, and the average ligand density (see Figure S7) rapidly decreases when moving away from their centre (the \enquote{hairy ball effect}). This facilitates {interlocking} of ligand bundles (see snapshot in Figure \ref{PMF}a), which reduces spacing as the core diameter decreases. In particular, we find that particles preferentially orient themselves so that an ordered bundle on one particle points into the groove between ordered bundles on the other particle, similar to how cogwheels fit together. This {bundle} interdigitation is distinct from the interdigitation of individual ligands that is often drawn in illustrations, and appears to be favored because it allows for a denser and more energetically stable packing of the ligands. In this case, it is the anisotropy of the ligand shell at low temperature, rather than only the particle curvature, that allows for particle separations that are less than twice the average shell thickness. As the core diameter increases, the ability of the bundles to interpenetrate decreases, resulting in particle separations that increase with core size in the ligand-dominated regime. On the other hand, as we enter the core-dominated regime, the strong vdW attraction between the metal cores exerts an increasing pressure on the disordered shell, which causes compression and leads to the \emph{decreased} spacing for the largest particles.

In summary, we propose that two transitions dominate the colloidal agglomeration of apolar nanoparticles with short ligands: a phase transition between ligand order and disorder (that can be driven by temperature or change in solvent quality), and a transition between core- and shell-dominated interaction (that depends on the core size and material). In the following, we quantitatively discuss the temperature- and size-dependant interaction potential between apolar nanoparticles using simulation results and a common analytical model.

Using constrained molecular dynamics simulations, we calculated the interaction potential between a pair of \SI{4}{\nano\meter} Au particles in decane at \(T_\mathrm{agglo} + \SI{20}{\celsius}, T_\mathrm{agglo} + \SI{5}{\celsius} \) and \(T_\mathrm{agglo} - \SI{10}{\celsius}\) (Figure \ref{PMF}b-d). The total potential curves (green triangles) show that the interaction between the particles changes from repulsive to attractive as the ligands align with one another and form ordered bundles on the surface of the particles. This dramatic change in how the particles interact with one another is due to a subtle change in the balance between the ligand-ligand and ligand-solvent interactions as the ligands order, with the former gradually becoming more attractive, especially at shorter range. In contrast, the vdW interaction between the Au cores (black squares) remains negligibly small until separations well below those observed experimentally. While accurate calculation of the interaction potential is difficult due to statistical sampling issues, we note that the minimum in the potential at \(T_\mathrm{agglo} + \SI{5}{\celsius} = T_\mathrm{order}\) occurs very close to the experimentally observed core surface spacing of \SI{2.4}{\nano\meter}. These results constitute a direct demonstration that ordering of ligands on small nanoparticles induces attraction between the particles even in a good solvent. Previous simulation studies of the interaction between small gold nanoparticles in solvents were limited to the high-temperature regime where the particles repel one another.\cite{Vlugt2008,Jabes2014,Yadav2016}

{We stress that the present discussion is relevant to the interaction between nanoparticles in good solvents such as decane. In vacuum, by contrast, the interaction between the particles is attractive irrespective of ligand conformation (see Fig. S6). The strong attraction in vacuum is due to the much stronger vdW attraction between the ligands in the absence of solvent, as has been noted previously for other nanoparticles.\cite{Landman2004,Vlugt2008,Widmer-Cooper2014,Waltmann2017}}

The existing analytical models for the interaction between apolar NPs assume a radially uniform ligand density around the particles and do not consider the ability of the ligand shell to change its order and symmetry. In the following, we summarize the state of the art in analytical modelling and briefly introduce a model based on the work of Khan et. al.\cite{khan2009self} that was modified for the particles used here. Most theoretical approaches assume a linear superposition of core-core van der Waals attraction, entropic repulsion due to ligand compression, and the free energy of mixing of ligands and solvents.\cite{khan2009self, goubet2011forces} The van der Waals (vdW) interaction  \(G_{vdW}\) is usually described using the reduced Hamaker coefficient \(A\) for the inorganic core interacting through an organic ligand/solvent medium \cite{bargeman1972van, ohara1995crystallization, morrison2002colloidal} and a geometrical factor that depends on the rescaled spacing \(\tilde{s}\) (center-to-center distance divided by the core diameter of the particles): 

\begin{equation}
     G_\mathrm{vdW}= -\frac{A}{12} \left( \frac{1}{\tilde{s}^2-1}+\frac{1}{\tilde{s}^2}+2 \ln \left( 1-\frac{1}{\tilde{s}^2}\right) \right)
     \label{equation:vdW}
\end{equation}

When the ligand shells overlap at a surface separation below one ligand length \(L\), ligand compression, and associated loss of conformational entropy, causes a repulsive force. This interaction \(G_\mathrm{com}\) is usually taken to be proportional to the ligand surface coverage \(\upsilon\),

\begin{equation}
    \frac{G_\mathrm{com}}{\mathrm{k_B} T}= \pi \upsilon d^2 \left(  (\tilde{s}-1)\left(ln\frac{\tilde{s}-1}{\tilde{L}}-1 \right) + \tilde{L}  \right),
\end{equation}

where \(d\) is the core diameter, \(\tilde{L}\) is the rescaled ligand length (ligand length divided by core diameter), \(\mathrm{k_B}\) Boltzmann's constant, and \(T\) the absolute temperature. 

Interactions between the solvent and the ligand shell add a free energy of mixing \(G_\mathrm{mix}\) that is often estimated using Flory-Huggins theory and can be either attractive or repulsive. When the particles are close enough to interpenetrate but do not deform (\(1+\tilde{L} < \tilde{s} < 1+2\tilde{L}\)), this interaction can be estimated as 

\begin{equation}
    \frac{G_\mathrm{mix1}}{k_bT}= \frac{\pi d^3}{2\nu_S} \phi^2 \left( \frac{1}{2}- \chi \right) \left( \tilde{s}- 1 -2\tilde{L} \right)^2
    \label{Flory1}
\end{equation}

When the particles are so close that the ligand shells are compressed (\(\tilde{s} < 1+\tilde{L}\)), the contribution becomes

\begin{equation}
    \frac{G_\mathrm{mix2}}{k_bT}= \frac{\pi d^3}{\nu_S} \phi^2 \tilde{L}^2 \left( \frac{1}{2}- \chi \right) \left( 3ln \frac{\tilde{L}}{\tilde{s}-1} + 2\frac{\tilde{s}-1}{\tilde{L}} - \frac{3}{2} \right)
    \label{Flory2}
\end{equation}

where \(\phi\ = \left( N_L \frac{\nu_L}{V_\mathrm{Sh}} \right)^2\) is the volume fraction occupied by the ligand shell, \(\nu_S\) is the volume of a solvent molecule,  \(\nu_L\) the volume of a ligand molecule, \(N_L\) the number of ligands per NP, \(V_\mathrm{Sh}\) the volume of the ligand shell, and \(\chi\) the Flory parameter that describes how well a single ligand molecule is solvated. This empirical parameter can be calculated from the Hildebrand solubility parameters of the solvent \(\delta_S\) and the ligand \(\delta_L\):

\begin{equation}
    \chi = \frac{V_S}{RT} (\delta_L- \delta_S)^2 + 0.34
\end{equation}

where \(V_S\) is the molar volume of the solvent and \(R\) the universal gas constant. A Flory parameter below \(0.5\) indicates that \(G_{mix} < 0\). {Note that Flory-Huggins theory has been developed for flexible polymers, not short ligands, and its use for ligand shells (as introduced by Khan in reference \cite{khan2009self}) should be seen merely as a phenomenological approach.}

We used this model to calculate interaction potentials (parameters are shown in table \ref{table:calculation}) for our particles at their agglomeration temperatures (Figure \ref{analytical}a). The predicted particle separations are larger than experimentally observed for all of the particles, and the predicted minima are too shallow to explain the observed agglomeration for all but the largest particles. While the attractive interactions between the gold cores (Figure \ref{analytical}b) are probably correctly represented in the model, it does not correctly describe the interaction of the ligand shells with the solvent and with other particles during agglomeration.

The simulations above indicate that much of this failure of the analytical model stems from its inability to account for ordering of the ligand shell. The ligand ordering has two important consequences: first, it provides an additional source of attraction between the particles that drives {agglomeration} in the absence of strong vdW attraction between the cores; and second, it reduces the compressive repulsion between the particles when they are oriented such that their ligand bundles can interdigitate, which allows for smaller particle separations than would otherwise be possible. The analytical model also appears to overestimate the energy required to compress the disordered ligand shell, predicting separation distances that are too large even in the core-dominated regime.

We conclude that the existing analytical models are not suitable to predict the stability of apolar nanoparticles, and that they predict incorrect particle spacings, regardless of the shell structure. An extension of the models would have to consider the temperature-dependant, possibly anisotropic structure of the shell. It is insufficient to merely change the solubility parameters.

The size and temperature dependence of particle stability has several interesting consequences. First, our results indicate that it might be possible to size-separate sufficiently large particles by temperature-induced agglomeration, because of the strong variation in agglomeration temperature with particle size. For example, the slope of the agglomeration curves was smallest for our largest particles (Figure \ref{Figure_1}b), despite all of our samples having similar size dispersity (Table \ref{table_size}). Second, the crystallinity of the agglomerates increased with increasing temperature for particles with diameters above \SI{7.5}{\nano\meter}, \textit{\textit{i.e.}} in or near the core-dominated regime. The transition is illustrated in Figure S8. This size-dependence may reflect the change in shape of the ligand shell from anisotropic to isotropic at \(T_\mathrm{agglo}\), although further work is needed to confirm this. We note that a similar increase in crystallinity with temperature has been reported for \SI{6.4}{\nano\meter} diameter Au particles where the {agglomeration} was driven by the addition of a poor solvent \cite{Born2012}.

\section{Conclusions}

We systematically evaluated the size-dependant stability of gold NPs with hexadecanethiol shells in decane experimentally and found a non-linear size dependency. Molecular dynamics simulations and analytical calculations of the core-core vdW interaction indicate that a transition between core- and shell-dominated agglomeration occurs around a core diameter of \SI{8}{\nano\meter}. This transition affects stability and particle spacing in the agglomerates, and can potentially occur for all hybrid particles consisting of a core that has a different Hamaker constant from the shell and a shell that is composed of discrete molecules that are anchored at one end. The transition diameter depends on the particular combination of core and ligand. 

As a rough estimate for whether a specific particle will be in the core- or the shell-dominated regime, one may calculate Equation \ref{equation:vdW} at \(\tilde{s} = 1+2\tilde{L}\), \textit{i.e.} the vdW attraction between the cores at the point where the ligands shells first come into contact. For our particles, this quantity corresponds to the red symbols in Figure \ref{analytical}b, which show that if \(G_\mathrm{vdW} > \SI{0.35}{k_BT}\) at this point then the particles are in the core-dominated regime and otherwise they are in the shell-dominated regime.

The dispersion used here arguably contains the simplest apolar nanoparticle available: A core with uniform density and a narrow size distribution is coated with a relatively uniform density of linear ligand molecules in a simple solvent that is structurally similar to the ligand. The fact that the existing theories are insufficient to predict its stability suggest that extended models are needed. Colloidal stability in the ligand-dominated regime depends so strongly on the difference in free energy between the ordered and disordered states that even small changes in ligand surface coverage should considerably affect it,\cite{Widmer-Cooper2016} which may explain the commonly observed batch-to-batch variations in the stability of freshly prepared apolar nanoparticles and their ageing \cite{lacava2015ageing, tim1996ageing}. {Other factors that are likely to affect the free energy balance of the ligand shell and associated stability of the particles in solvents include more complex ligand structures (for example, a kink in the case of oleylamine), mixtures of ligands, and lateral diffusion of ligands on the nanocrystal surface \cite{Boles2016, yang2016entropic}.}

The results presented here suggest that even small changes in ligand and solvent length may have a strong effect on colloidal stability. Initial experiments do indeed indicate ligand and solvent dependencies that run counter to the predictions of conventional colloid theory. These results will be presented in forthcoming publications.

\section{Methods}

All chemicals were obtained from Sigma Aldrich (unless noted otherwise) and used without further purification. 

\subsection{Nanoparticle synthesis} 

Gold nanoparticles (Au NPs) with diameters between \SI{4}{\nano\meter} and \SI{9.7}{\nano\meter} were synthesized using a modified protocol based on Wu and Zheng \cite{Zheng2013}. Au NPs with a diameter of \SI{8}{\nano\meter} were produced as follows. A mixture of \SI{8}{\milli\liter} benzene (puriss. $\ge$ 99.7\%), \SI{8}{\milli\liter} oleylamine (technical grade, 70\%) and \SI{100}{\milli\gram} of HAuCl\(_4\)x H\(_2\)O was stirred at \SI{20}{\celsius} and \SI{500}{\radian\per\minute} for \SI{1}{\minute} under argon atmosphere. Afterwards \SI{40}{\milli\gram} tert-butylamine borane (ABCR, 97\%) which was dissolved in \SI{2}{\milli\liter} benzene and \SI{2}{\milli\liter} oleylamine (OAm) was added to the solution. The color of the solution immediately became dark purple. After stirring for \SI{60}{\minute} at \SI{20}{\celsius}, the nanoparticles were purified once by precipitating with \SI{30}{\milli\liter} ethanol and centrifugation at \SI{4000}{\radian\per\minute} for \SI{5}{\minute}. The precipitated nanoparticles were redispersed in \SI{20}{\milli\liter} heptane (puriss. $\ge$ 99\%). AuNP with a diameter of \SI{4}{\nano\meter} were obtained using pentane (reagent grade, 98\%) instead of benzene and stirring for \SI{30}{\minute} before adding the tert-butylamine borane complex.

\subsection{Nanoparticle characterization}

The core size of the NPs was measured by small angle X-ray scattering (SAXS, device explained in more details below) and by analyzing transmission electron microscopy (TEM) micrographs. The angle-dependant scattering intensity was then fitted using SASfit (Version 0.94.6) provided by the Paul Scherrer Institute. TEM micrographs were taken with a JEOL JEM 2010 at \SI{200}{\kilo\volt}. Around 2000 particles were counted with the ImageJ 1.45s software for each size, and the arithmetic mean and the standard deviation were calculated.

\begin{table}[ht]
\centering
\small
  \caption{AuNPs used for this study, with diameters obtained from transmission electron microscopy and small angle X-ray scattering.}
  \label{table_size}
  \begin{tabular*}{0.8\textwidth}{@{\extracolsep{\fill}}cccc}
    \hline
    Number & d (TEM) & d (SAXS)\\
    \hline
    Au01 & 4.1 nm $\pm$ \SI{10}{\percent} & 4.1 nm $\pm$ \SI{9.3}{\percent}\\
    Au02 & 5.6 nm $\pm$ \SI{8.4}{\percent} & 5.5 nm $\pm$ \SI{8.3}{\percent}\\
    Au03 & 6.4 nm $\pm$ \SI{6.3}{\percent} & 6.2 nm $\pm$ \SI{8.3}{\percent}\\
    Au04 & 7.4 nm $\pm$ \SI{7.4}{\percent} & 7.5 nm $\pm$ \SI{6.7}{\percent}\\
    Au05 & 8.5 nm $\pm$ \SI{7.1}{\percent} & 8.3 nm $\pm$ \SI{6.7}{\percent}\\
    Au06 & 8.9 nm $\pm$ \SI{8.5}{\percent} & 8.9 nm $\pm$ \SI{6.8}{\percent}\\
    Au07 & 9.8 nm $\pm$ \SI{5.8}{\percent} & 9.7 nm $\pm$ \SI{7.3}{\percent}\\
    \hline
  \end{tabular*}
\end{table}

\subsection{Ligand exchange}

Ligand exchange was performed with previous published methods \cite{kister2016pressure}. Oleylamine-stabilized AuNP were then heated to \SI{80}{\celsius} under argon. 1-hexadecanethiol ($\ge$ 95.0\% GC, 10 times the molar amount of gold), was added to the solution and the mixture was stirred at \SI{80}{\celsius} for \SI{10}{\minute}. The resulting particles were purified once by precipitation with ethanol, centrifugation, and resuspension in decane ($\ge$ 95\%).

\subsection{Thermogravimetric analyses}

Thermogravimetric analyses were performed at a Netzsch STA 449 F3 Jupiter. The measurements started at room temperature and run until \SI{800}{\celsius}. The heating rate was kept at \SI{10}{\kelvin \per \minute}. All measurements were done under an inert atmosphere.

\subsection{Small-Angle X-ray Scattering}

All scattering experiments were performed at a Xeuss 2.0 from Xenocs SA (Grenoble, France). The setup was equipped with a copper $K_\alpha$ X-ray source with a wavelength $\lambda =$ \SI{0.154}{\nano\meter} and a  PILATUS 1M Hybrid Photon Counting detector from DECTRIS (Baden, Switzerland). The sample to detector distance was kept at \SI{1200}{\milli\meter}. 

A quantity of \SI{20}{\micro \litre} to \SI{40}{\micro\litre} of the respective dispersion was filled into single-used capillaries with a diameter of \SI{2}{\milli\meter} that was sealed with epoxy and introduced to the vacuum of the sample chamber. Pure solvent was measured in a reference capillary with the same diameter. 

Temperature was controlled using a Peltier-based temperature stage (Omega CN8200) in a range between \SI{-20}{\celsius} and \SI{120}{\celsius}. All measurements started at high temperatures to ensure that the NPs were deagglomerated. The temperature was then decreased or increased in steps of \SI{2}{\celsius} or \SI{5}{\celsius}. At each step, \SI{20}{\minute} equilibration time was allowed before acquisition of scattering data during \SI{10}{\minute}. 

To achieve the pure particle signal $I(q)_{NPs}$, the total scattering intensity $I(q)$ was corrected by the transmission factor of the NPs $T_{NPs}$, followed by a subtraction with the scattered intensity $I(q)_S$ of the pure solvent, which was also corrected by the transmission factor $T_S$ of the solvent \cite{schnablegger2013saxs}. This is shown by the equation below. The correction factors $T_{NPs}$ and $T_S$ were extracted from the pindiode signal which is mounted in the direct beam.

\begin{equation}
     I(q)_{NPs} = \frac{I(q)}{T_{NPs}} - \frac{I(q)_S}{T_S}
\end{equation}

To extract the degree of agglomeration \textit{versus} the temperature, the $"$structure-factor$"$ $S(q)$ was calculated by dividing all corrected spectra with the form-factor $P(q)_{NPs}$. The scattered intensity is a product of $S(q)$ and $P(q)$. When NPs are fully dispersed (high temperatures) $S(q)$ is equal one. Due to this the signal at high temperature is $P(q)$.

\begin{equation}
     S(q)_{NPs} = \frac{I(q)_{NPs}}{P(q)_{NPs}}
\end{equation}

\subsection{Molecular dynamics simulations}
The NPs were modeled as spherical Au cores covered in alkyl thiolate ligands in the presence of n-decane, as illustrated in Figure S9. The ligands were assumed to be irreversibly bound to the gold surface at the surface coverage determined by TGA measurements (\SI{5.5}{\nano\meter^{-2}}). The positions of the sulfur atoms were constrained by the RATTLE algorithm \cite{Andersen1983} and determined by placing them on a spherical shell around the implicit metallic core (\SI{0.15}{\nano\meter} further out), and allowing them to find their optimal positions on this shell subject to a repulsive interaction between them (standard Coulombic potential with a dielectric constant $\epsilon$ = \SI[mode=text]{10}{kcal.mol^{-1}}, truncated at \SI{24}{\angstrom}). This ensured that the binding sites were approximately equidistant from one another. The sulfur atoms were subsequently treated as part of the rigid core of the particle.
The rest of the ligand and solvent molecules were modeled using a united-atom representation, with each CH$_x$ group being represented by a single particle.
These particles interacted with one another according to the 12-6 Lennard-Jones (LJ) potential, with parameters as used and described previously \cite{Widmer-Cooper2014}. Bond stretching, bond bending, and dihedral torsion terms were also considered within each molecule \cite{Martin1998}. 
The interaction between the CH$_x$ groups and the Au core was modelled using a 9-3 LJ potential (with $\epsilon$/k$_b$ = \SI{88}{\kelvin} and $\sigma$ = \SI{3.54}{\angstrom},\cite{Pool2007} truncated at \SI{30}{\angstrom}). This provides a good approximation to CH$_x$-core interactions for NPs $\geq$ \SI{4}{\nano\meter} in diameter.

Molecular dynamics (MD) simulations on systems of up to 565,000 particles were performed using LAMMPS \cite{Plimpton1995} with periodic boundary conditions, at temperatures ranging from \SIrange{245}{330}{\kelvin} (depending on the core diameter).
Individual NPs were initially equilibrated in explicit decane at a temperature sufficiently high to ensure that the ligands were in the disordered state (e.g., \SI{400}{\kelvin}).
During this run, the periodic simulation cell was slowly compressed until the solvent density far from the NP was equal to the experimental density of pure decane at the relevant temperature.
Subsequent runs were carried out at fixed pressure (\SI[mode=text]{80}{atm}) and temperature, maintained with a Nos\'e-Hoover thermostat and barostat. This yielded bulk solvent densities within \SI{1}{\percent} of experimental values. Systems were equilibrated for at least \SI{12}{\nano\second} before \SIrange{1}{2}{\nano\second} production runs were performed.

\subsubsection{Interaction potentials}
The interaction potential between a pair of \SI{4}{\nano\meter} particles was calculated as a Potential of Mean Force (PMF) at selected temperatures using constrained MD.
As the particles were brought together (at a rate of \SI{1}{\angstrom\per\nano\second}), they were allowed to freely rotate about their centers of mass.
In order to allow the ligands to reorganize and find more stable configurations at and below \(T_\mathrm{order}\), we performed an additional thermal annealing step at separations where the ligand shells overlapped.
This was done by increasing the temperature of these systems by \SI{50}{\kelvin} over \SI{1}{\nano\second} and subsequently cooling it back to the initial temperature over the course of \SI{3}{\nano\second}.
We found that long subsequent simulation times ($>$\SI{10}{\nano\second} at each separation) were required to adequately sample the PMF, especially at lower temperatures where the ligands are less mobile.

The spherical gold cores were assumed to interact with each other via the Hamaker potential \cite{Hamaker1937}, with a Hamaker constant of \SI{2}{\electronvolt} \cite{Ederth2001}. This approach treats the solvent and ligands as a single continuum, with the interaction constant scaled to include the effect of the hydrocarbon medium.

The PMF between two nanoparticles is given by
\begin{equation}
    \phi_{MF}(r)=\int_{r}^{\infty} F_{mean}(s) ds
\end{equation}
Where $F_{mean}$ is the average force in the direction of the line connecting the two particles and is given by
\begin{equation}
    F_{mean}(r)=\frac{1}{2}\langle(\vec{F}_{2}-\vec{F}_{1})\cdot\vec{r}\rangle_{NVT}
\end{equation}
In the above, $\vec{F}$$_1$ and $\vec{F}$$_2$ are the total forces acting on the first and second NP, respectively, $\vec{r}$ is the unit vector pointing from one particle's center to the other's, and the angular brackets denote an average in the canonical ensemble.

\begin{table}[ht]
\centering
\small
  \caption{Parameters used for analytical calculations of the interaction potential using equations 1-5 \cite{khan2009self, kokkoli1999effect}}
  \label{table:calculation}
  \begin{tabular}{||c|c||} \hline
    \hline
    
   Hamaker \(A\) for gold & \SI{75.5}{k_BT} \\ \hline
   Ligand length \(L\) for hexadecanethiol & \SI{2.2}{\nano\meter} \\ \hline
   Ligand density \(\upsilon\) & \SI{5.5}{\nano\meter^{-2}} \\ \hline
   Volume solvent molecule \(\nu_S\) & \SI{0.324}{\nano\meter^3}
 \\ \hline
   Volume ligand molecule \(\nu_L\) & \SI{0.505
}{\nano\meter^3} \\ \hline
   Molar volume solvent \(V_S\) & \SI{0.000195}{\meter^3/mol} \\ \hline
   Solubility parameter \(\delta\) decane & \SI{1.58E4}{\sqrt Pa} \\ \hline
   Solubility parameter \(\delta\) hexadecane & \SI{1.63E4}{\sqrt Pa} \\ \hline
   Solubility parameter \(\delta\) hexadecanethiol & \SI{1.69E4}{\sqrt Pa} \\ \hline

  \end{tabular}
\end{table}

\acknowledgement
A.W. thanks the Australian Research Council for a Future Fellowship (FT140101061). D.M. was supported by the ARC Centre of Excellence in Exciton Science (CE170100026). Computational resources were generously provided by the University of Sydney HPC service, the National Computational Infrastructure National Facility (NCI-NF) Flagship program and the Pawsey Supercomputer Centre Energy and Resources Merit Allocation Scheme. T.K. and T.K. thank the DFG – Deutsche Forschungsgemeinschaft for funding, and E. Arzt for his continuing support of this project.

\suppinfo

Supporting Information shows SAXS and TEM data of the oleylamine capped AuNPs, TGA data of 1-hexadecanethiol capped AuNPs, analyzed SAXS data of AuNPs, average dihedral angles in vacuum and radial density distributions of ligand and solvent (from simulation), structure factors of AuNPs and a snapshot of the simulation setup.

\bibliography{bib}

\end{document}

\begin{figure}[ht]
    \centering
    \includegraphics[width=0.7\linewidth]{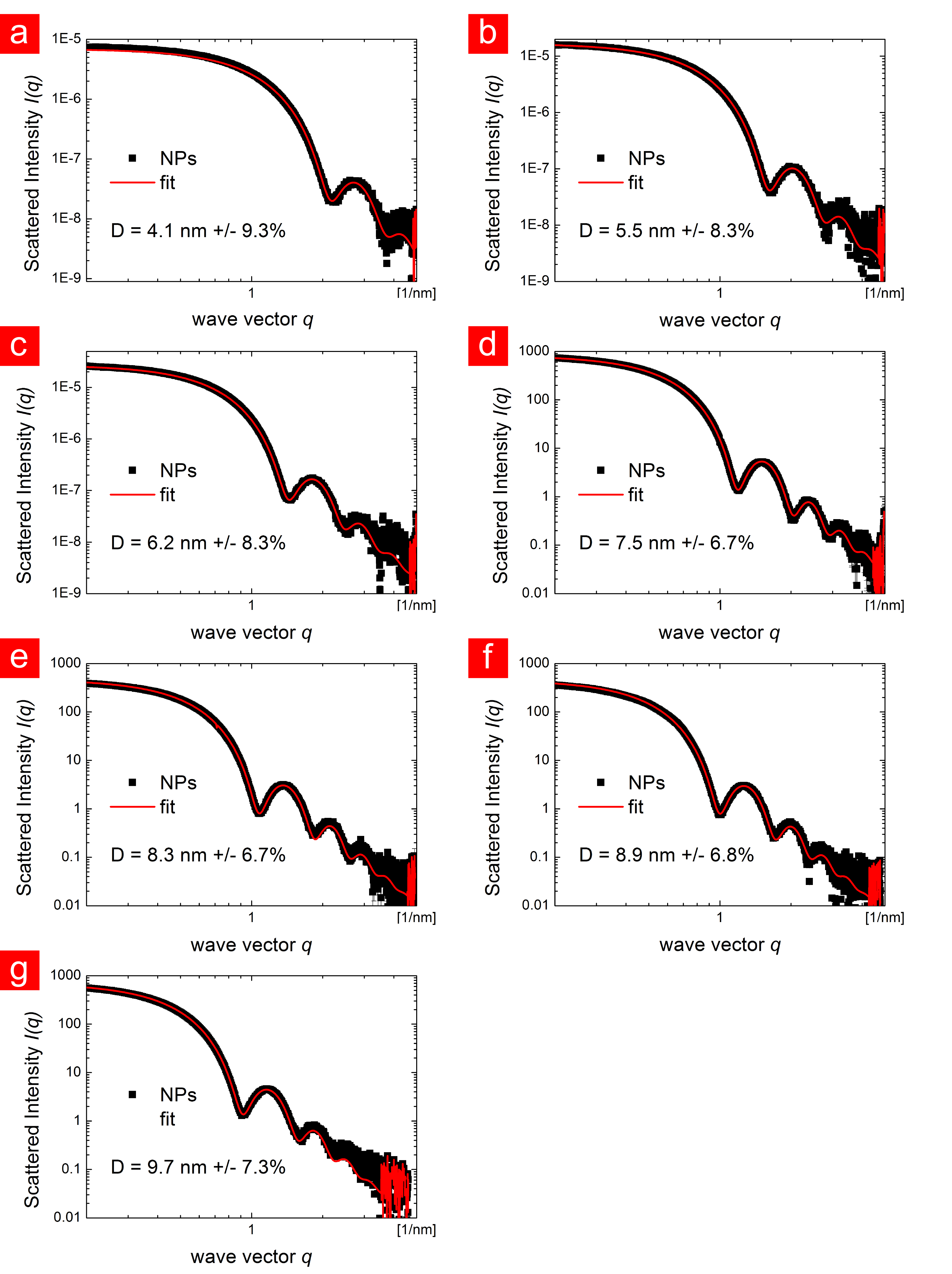}
    \caption{Small angle X-ray scattering data of the used AuNPs. The data was fitted to a sphere scattering model to obtain diameter and width of size distribution.} 
    \label{Size_SAXS}
    \vspace{0.5em}
\end{figure}
\begin{figure}[ht]
    \centering
    \includegraphics[width=0.5\linewidth]{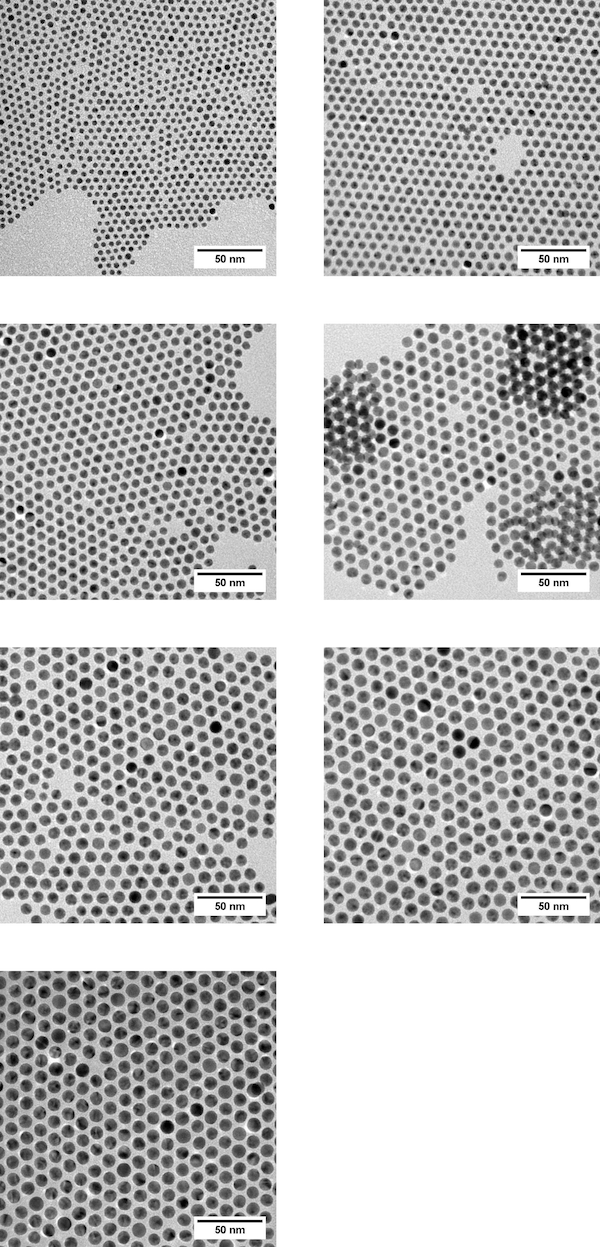}
    \caption{Transmission electron microscopy micrographs of the used AuNPs. Image analysis was performed to obtain diameter and width of size distribution.} 
    \label{Size_TEM}
    \vspace{0.5em}
\end{figure}

Figures \ref{Size_SAXS} and \ref{Size_TEM} show SAXS and TEM data of the as-synthesized, oleylamine-capped AuNPs. We used the program SASfit 0.94.6 from the Paul Scherrer Institute to fit the SAXS data with a sphere scattering model and ImageJ 1.45s to evaluate the particle size from TEM micrographs.

\begin{figure}[ht]
    \centering
    \includegraphics[width=0.45\linewidth]{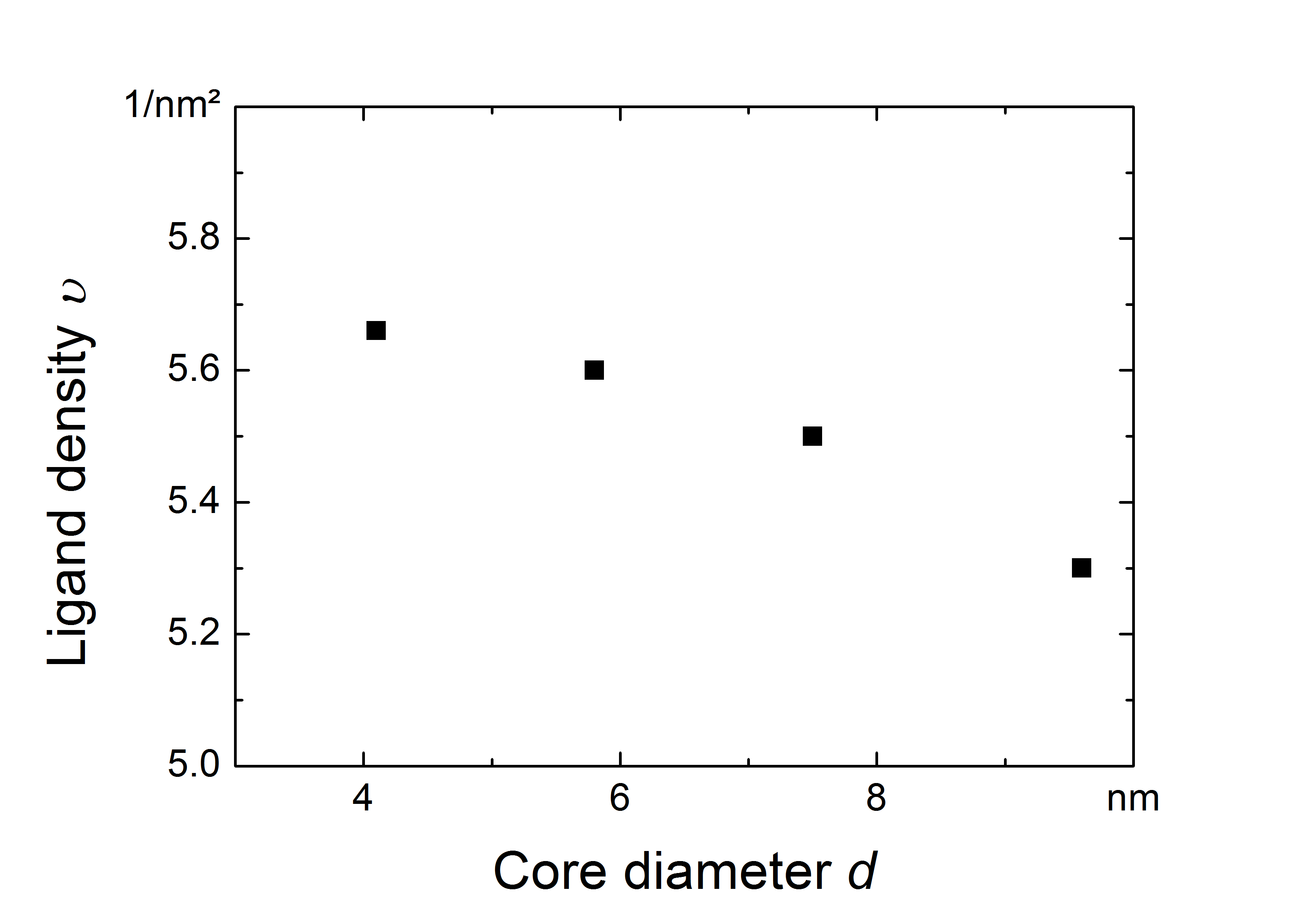}
    \caption{Thermogravimetric analysis data of AuNPs with 4 different sizes. The ligand density was between \SI{5.65}{\nano\meter^{-2}} and \SI{5.3}{\nano\meter^{-2}} for all particles.} 
    \label{SI_TGA}
    \vspace{0.5em}
\end{figure}

Figure \ref{SI_TGA} shows the evaluated TGA data for 4 different core diameters. After ligand exchange with 1-hexadecanethiol, the AuNPs were washed 3 to 5 times by precipitation with methanol/ethanol and centrifugation and finally redispersed in hexane. The dispersion was dried and \SIrange{20}{40}{\milli\gram} of AuNPs were placed in TGA crucibles for each measurement.

\begin{figure}[ht]
    \centering
    \includegraphics[width=0.8\linewidth]{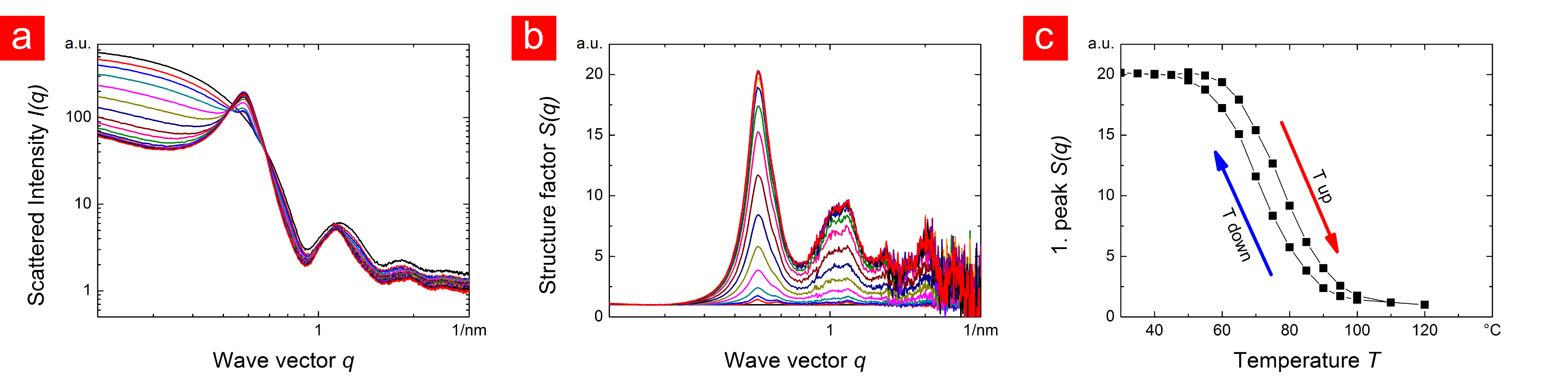}
    \caption{Temperature-dependent scattering of AuNPs with a diameter of \SI{9.7}{\nano\meter}. a) Raw data. b) The calculated structure factors. The peaks increase upon cooling. c) Evolution of the height of the first peak during a temperature cycle from \SI{120}{\celsius} to \SI{30}{\celsius} and back to \SI{120}{\celsius} over \SI{16}{\hour}.} 
    \label{SI_data}
    \vspace{0.5em}
\end{figure}

\begin{figure}[ht]
    \centering
    \includegraphics[width=0.45\linewidth]{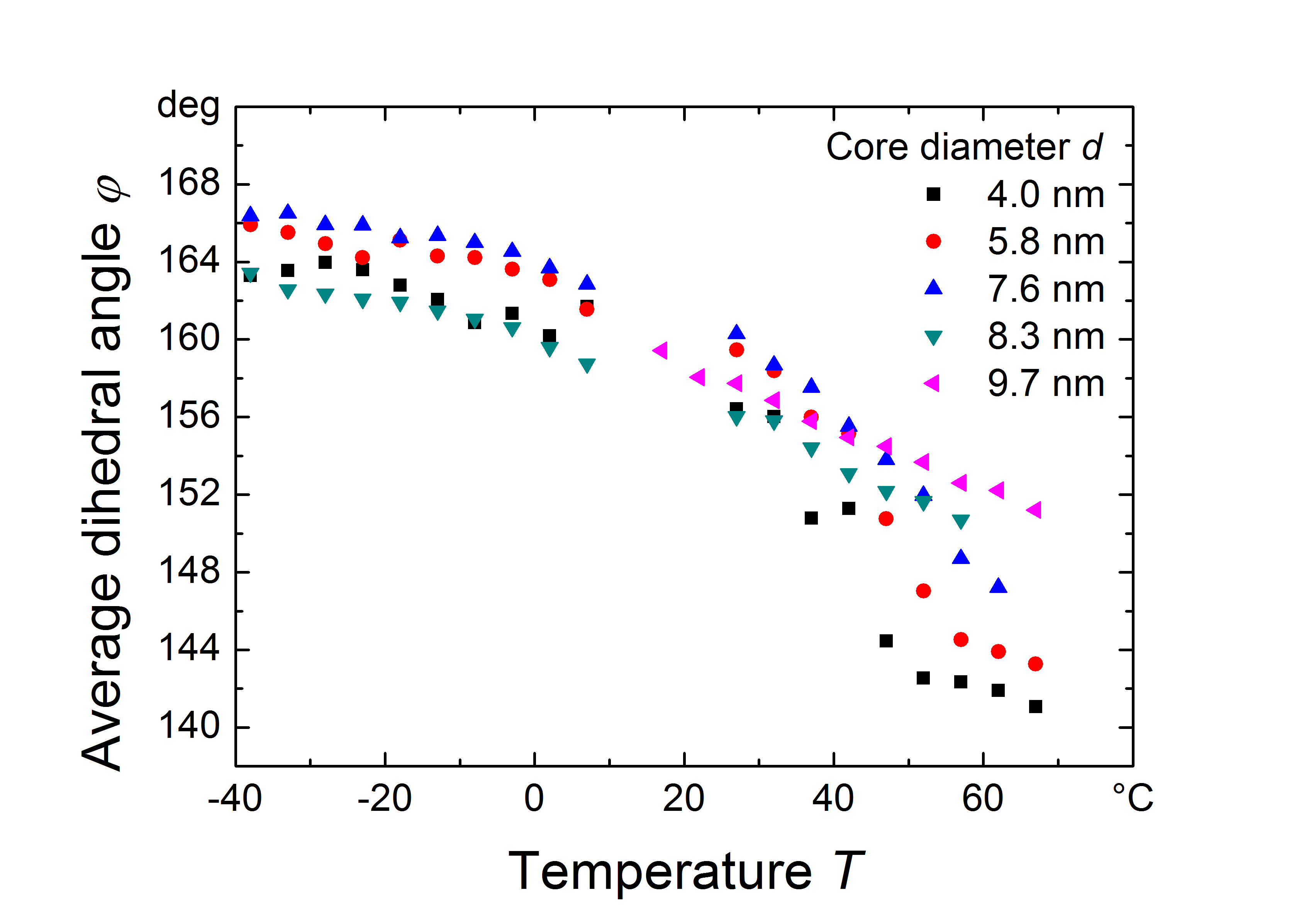}
    \caption{Average dihedral angle of the ligands as a function of temperature for different particle core diameters in vacuum. As in decane, there is a crossover in behavior near a diameter of \SI{8}{\nano\meter}, but this is no longer associated with a large drop in the ordering temperature.} 
    \label{SI_vacuum}
    \vspace{0.5em}
\end{figure}

\begin{figure}[ht]
    \centering
    \includegraphics[width=0.8\linewidth]{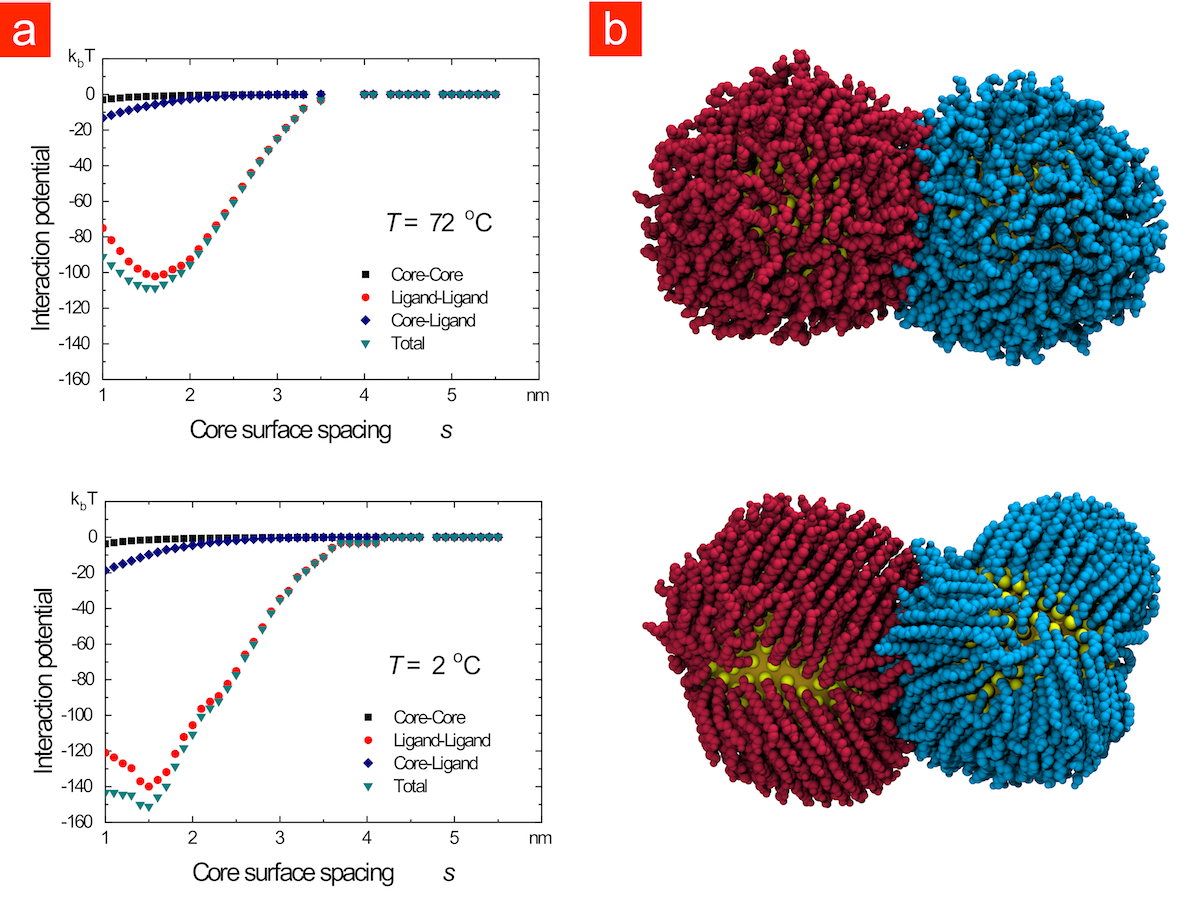}
    \caption{{(a) Interaction potentials for pairs of \SI{4}{\nano\meter} Au particles in vacuum obtained by calculating the potential of mean force between the particles using constrained molecular dynamics simulations. The total potential is green, the ligand-ligand contribution is red, the ligand-core contribution is blue, and the core-core contribution is black. The interaction potential in vacuum is attractive irrespective of the conformation of the ligands. (b) Simulation snapshots for the two particles at \SI{1.5}{\nano\meter} separation at the respective temperatures.}} 
    \label{SI_PMF_vacuum}
    \vspace{0.5em}
\end{figure}

\begin{figure}[ht]
    \centering
    \includegraphics[width=0.8\linewidth]{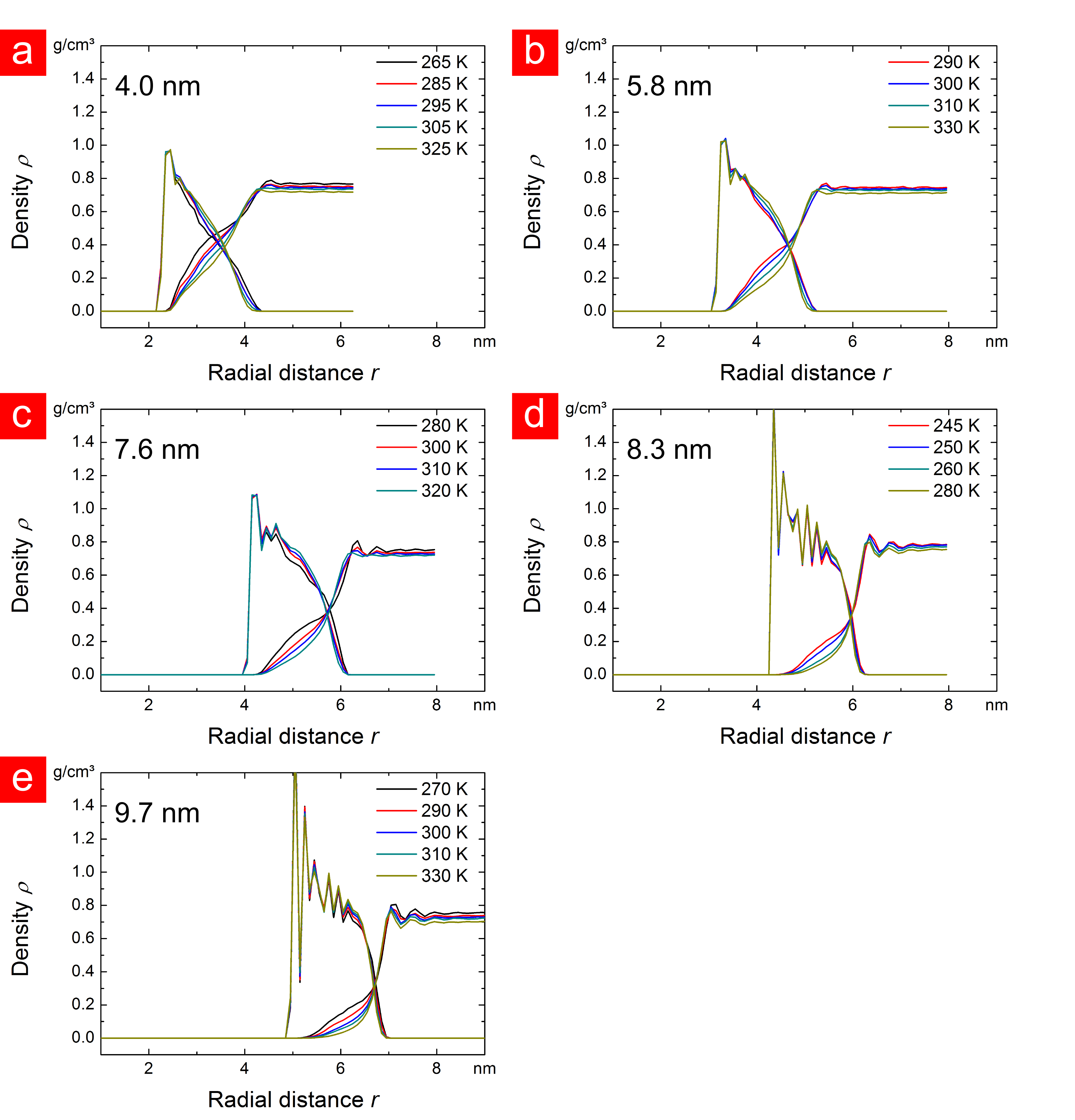}
    \caption{Radial density distributions for the ligand and solvent for different particle core diameters, plotted as a function of the distance \(r\) from the center of the nanoparticle core. The dark blue lines indicate the density profile at the ligand ordering temperature \(T_\mathrm{order}\) for each particle. For all core sizes, as the temperature decreases and the ligands order, the space between bundles on the surface of the particle becomes larger, allowing for more solvent to occupy that region. As the particles get larger, this space becomes smaller and the interface between the ligands and the solvent becomes more sharply defined, limiting the amount of solvent that is able to enter the ligand shell.}
    \label{SI_density}
    \vspace{0.5em}
\end{figure}

\begin{figure}[ht]
    \centering
    \includegraphics[width=0.8\linewidth]{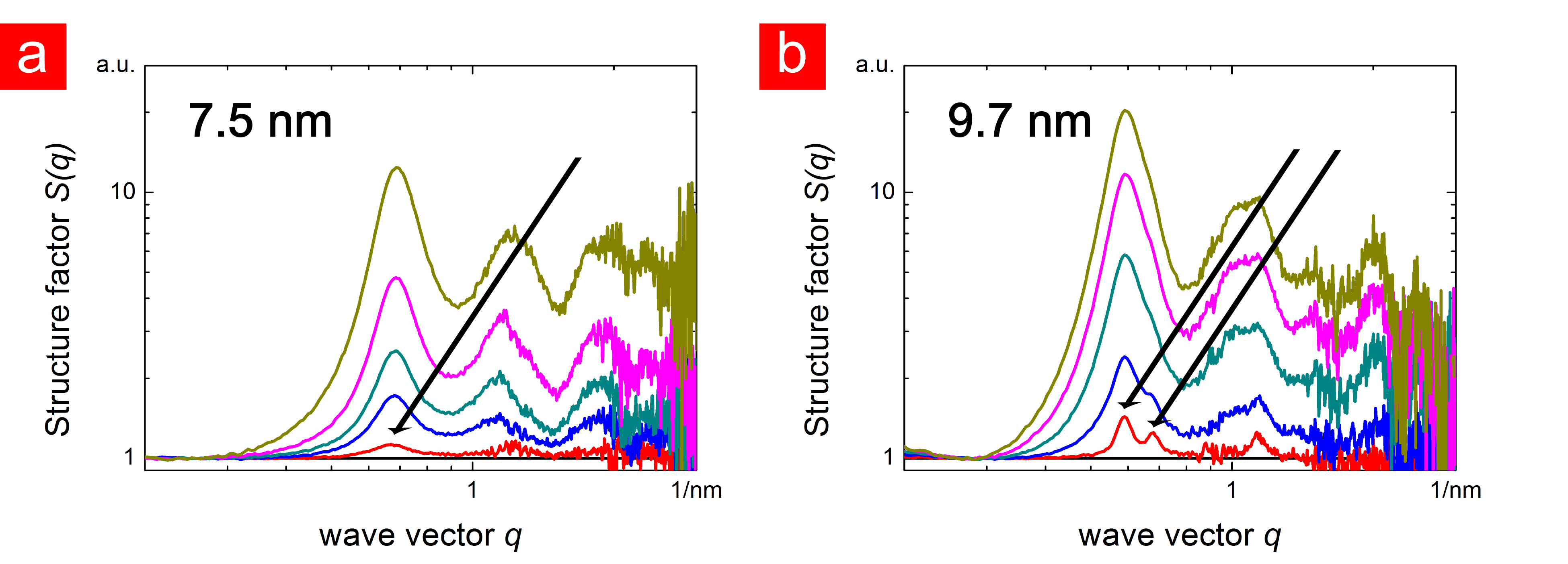}
    \caption{Temperature dependent structure factor for AuNPs with diameters of \SI{7.5}{\nano\meter} and \SI{9.7}{\nano\meter}. Smaller NPs formed amorphous agglomerates (no visible crystalline structure at the beginning of the agglomeration). For bigger NPs crystalline structure (fcc) appeared at the beginning of the agglomeration. This is indicated by the formation of two peaks (marked by arrows).} 
    \label{SI_structure}
    \vspace{0.5em}
\end{figure}

\begin{figure}[ht]
    \centering
    \includegraphics[width=0.45\linewidth]{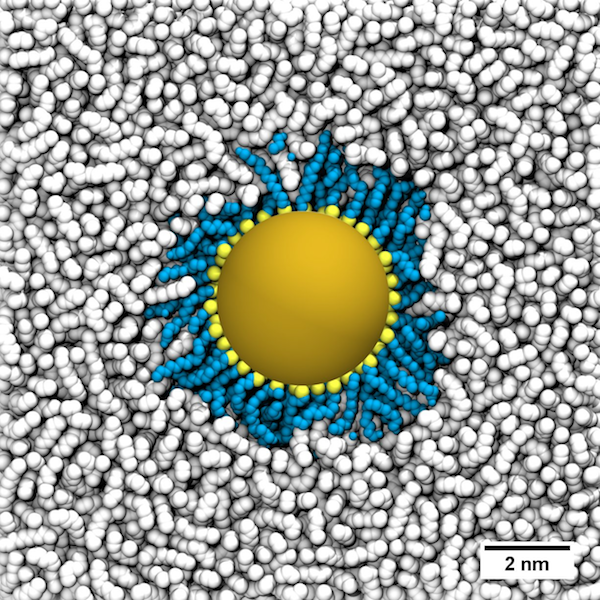}
    \caption{Snapshot of a simulation cell that has been cut in half showing the set up of our model for a gold nanocrystal (golden sphere at the center) covered in hexadecyl thiolate ligands (yellow and blue particles) and solvated by n-decane (white particles).}
    \label{SI_ligand_vs_core}
    \vspace{0.5em}
\end{figure}